\documentstyle[aps,prl]{revtex}
\input{epsf}
\begin{document}
\title{Statistical mechanical aspects  of joint source-channel coding}
\author{Ido Kanter and Haggai Kfir}
\address{
Minerva Center and the Department of Physics, Bar-Ilan University,
Ramat-Gan 52900, Israel}
\maketitle
\begin{abstract} 
An MN-Gallager Code over Galois fields, $q$, based on the Dynamical
Block Posterior probabilities (DBP) for messages with a given set of
autocorrelations is presented with the following main results: (a) for
a binary symmetric channel the threshold, $f_c$, is extrapolated for
infinite messages using the scaling relation for the median
convergence time, $t_{med} \propto 1/(f_c-f)$; (b) a degradation in
the threshold is observed as the correlations are enhanced; (c) for a
given set of autocorrelations the performance is enhanced as $q$ is
increased; (d) the efficiency of the DBP joint source-channel coding
is slightly better than the standard gzip compression method; (e) for
a given entropy, the performance of the DBP algorithm is a function of
the decay of the correlation function over large distances.
\end{abstract}

With the rapid growth of information content in today's wire and
wireless communication, there is an increasing demand for efficient
transmission systems.  A significant gain in the transmission
performance can be achieved by the application of the joint
source-channel coding technique, which has attracted much attention
during the recent past, see for instance
\cite{shamail1,shamail2,shamail3,shamail4,shamail5,shamail6}.  Roughly
speaking, source coding is mainly a data compression process that aims
at removing as much redundancy as possible from the source signal,
whereas channel coding is the process of intelligent redundancy
insertion so as to be robust against channel noise. These two
processes, source coding and channel coding, seem to act in
opposition, where the first/second process shrinks/expands the
transmitted data.  For illustration, assume that our compression {\it
shrinks} the size of the source signal by a factor $2$ and in order to
be robust against channel noise we have to {\it expend} our compressed
file by a factor $4$. Hence, the length of the transmitted sequence is only
twice the length of the uncompressed source.

The source-channel coding theorem of Shannon\cite{Shannon-48}
indicates that if the minimal achievable source coding rate of a given
source is below the capacity of the channel, then the source can be
reliably transmitted through the channel, assuming an infinite source
sequence. This theorem implies that source coding and channel coding
can be treated {\it separately} without any loss of overall
performance, hence they are fundamentally separable. 
Practically, the
source can be first efficiently compressed and then an efficient error
correction method can be used.

The objective of joint source-channel coding is to combine both source
(compression) and channel (error correction) coding into one mechanism
in order to reduce the overall complexity of the communication while
maintaining satisfactory performance.  Another possible advantage of
the joint source-channel coding is the reduction of the sensitivity to
a bit error in a compressed message.

In a recent paper\cite{ido-new} a particular scheme based on
a statistical mechanical approach for the implementation of the joint 
source-channel coding was presented and the main steps are briefly
summarized below.  The original boolean source is first mapped to
a binary source\cite{sourlas,sourlas1} $\left\{ x_{i}\pm1\right\} ~i=1,...,L$,
and is characterized by a finite set, $k_0$, of autocorrelations
\begin{equation}
C_{k}=\frac{1}{L}\sum _{i=1}^{L}x_{i}x_{\left(i+k\right)\:
\mathbf{mod}\: L}
\end{equation}
\noindent where $k \le k_0$ is the highest autocorrelation taken.  The
number of  sequences oeying these $k_0$ constraints is
given by
\begin{equation}
\Omega = Tr \prod_{k=1}^{k_0} \delta \left(\sum _{j}x_{j}x_{j+k}-
C_{k}L\right)
\end{equation}
Introducing the Fourier representation of the delta functions and then
using the standard transfer matrix (of size $2^{k_0} \times
2^{k_0}$) method,\cite{baxter} one finds $\Omega=\int
dy_k\exp\{-L\lbrack \sum y_{k}C_{k} - \ln \lambda_{max}(\{ y_{k}\})
\rbrack \}$, where $\lambda_{max}$ is the maximal eigenvalue of the
corresponding transfer matrix.  For large $L$, using the saddle point
method the entropy, $H_2(\{C_k\})$, is given in the leading order by
\begin{equation}
H_2\left(\{C_k\}\right)=\frac{\frac{1}{k_0}\ln \lambda_{max}\left
(\{ y_{k}\}\right)-\sum_{i=1}^{k_0}y_{k}C_{k}}
{\ln 2}\label{entropy-ck}
\end{equation}
\noindent where $\{y_{k}\}$ are determined from the
saddle point equations of $\Omega$.\cite{ido-new} Assuming Binary
Symmetric Channel (BSC) and using Shannon's lower bound, the channel
capacity of sequences with a given set of $k_0$ autocorrelations is
given by
\begin{equation}
C=\frac{1-H_{2}\left(f\right)}{H_{2}(\left\{C_{k}\}\right)-
H_{2}\left(P_{b}\right)}
\end{equation}
\noindent where $f$ is the channel bit error rate and $p_b$ is a bit
error rate.  The saddle point solutions derived from eq. 3 indicate
that the equilibrium properties of the one dimensional Ising spin
system
\begin{equation}
H=-\sum_i \sum_{k=1}^{k_0} \frac{y_{k}}{\beta }
x_{i}x_{i+k}
\end{equation}
\noindent obey in the leading order the autocorrelation constraints of
eq. 2.  Note that in the typical scenario of statistical mechanics,
one of the main goals is to calculate the partition function and the
equilibrium properties of a {\it given Hamiltonian}.  Here we present
a prescription of how to solve the reverse question. Given the desired
macroscopic properties, the set of the autocorrelations, the goal is
to find the {\it appropriate Hamiltonian} obeying these macroscopic
constraints.  This property of the effective Hamiltonian, eq. 5, is
used in simulations below to generate an ensemble of signals (source
messages) with the desired set of autocorrelations.

The decoding of symbols of $k_0$ successive bits is based on the
standard message passing introduced for the MN decoder over Galois
fields with $q=2^{k_0}$\cite{LDPC-GF(q)} and with the following
modification. The horizontal pass is left unchanged, {\it but a
dynamical set of probabilities assigned for each block are used in the
vertical pass}. The Dynamical Block Probabilities (DBP), $\{P_n^c\}$,
are determined following the current belief regarding the neighboring
blocks and are given by
\begin{eqnarray}
\gamma _{n}^{c} & = & S_{I}\left(c\right)\left(\sum
_{l=1}^{q}q_{L}^{l}S_{L}\left(l,c\right)\right)\left(\sum
_{r=1}^{q}q_{R}^{r}S_{R}\left(c,r\right)\right)\nonumber \\ P_{n}^{c}
& = & \frac{\gamma _{n}^{c}}{\sum _{j=1}^{q}\gamma
_{n}^{j}}\label{tm-vertical-pass}
\end{eqnarray}
\noindent where $l/r/c$ denotes the state of the left/right/center
($n\!-\!1\,/\,n\!+\!1\,/\,n$) block respectively and
$q_{L}^{l}/q_{R}^{r}$ are their posterior
probabilities. $S_I(c)=e^{-\beta H_I}$, where $H_I$ is the inner
energy of a block of $k_0$ spins at a state $c$, see eq. 5.  Similarly
$S_L(l,c)$ ($S_R(c,r)$) stands for the Gibbs factor of consecutive
Left/Center (Center/Right) blocks at a state $l,c$
$(c,r)$.\cite{ido-new}

Note that the complexity of the calculation of the block prior
probabilities is $O(Lq^2/ \log q)$ where $L/\log q$ is the number of
blocks.  The decoder complexity per iteration of the MN codes over a
finite field $q$ can be reduced to order
$O(Lqu)$\cite{David_Mackay,David_Mackay1,David_Mackay2}, where $u$
stands for the average number of checks per block.  Hence the total
complexity of the DBP decoder is of the order of $O(Lqu+Lq^2/ \log
q)$.

In this Letter we examine the efficiency of the DBP-MN decoder as a
function of the maximal correlation length taken, $k_0$, the strength
of the correlations, the size of the finite fields $q$, and we compare
this efficiency with the standard gzip compression procedure.  A
direct answer to the above questions is to implement exhaustive
simulations on increasing message length, various finite fields
$q$, and sets of autocorrelations, which result in the bit error
probability versus the flip rate $f$. Besides the enormous
computational time required , the conclusions would be controversial
since it is unclear how to compare, for instance, the performance as a
function of $q$; with the same number of transmitted blocks or with
the same number of transmitted bits.

In order to overcome these difficulties, for a given MN-DBP code over
GF(q) and a set of autocorrelations, the threshold $f_c$ is estimated
from the scaling argument of the convergence time, which was
previously observed for $q=2$\cite{KS-LDPC,KS-Gaussian}.  The median
number of message passing steps, $t_{med}$, necessary for the
convergence of the MN-DBP algorithm is assumed to diverge as the level
of noise approaches $f_c$ from below. More precisely, we found
that the scaling for the divergence of $t_{med}$ is independent of $q$
and is consistent with
\begin{equation}
t_{med} = {A \over f_c-f}
\end{equation}
\noindent where for a given set of autocorrelations and $q$, $A$ is a
constant. Moreover, for a given set of autocorrelations and a finite
field $q$, the extrapolated threshold $f_c$ is independent of $L$, as
demonstrated in Fig. 1.  This observation is essential to determine
the threshold of a code based on the above scaling behavior. Note that
the estimation of $t_{med}$ is a simple computational task in
comparison with the estimation of low bit error probabilities for
large $L$, especially close to the threshold. We also note that the
analysis is based on $t_{med}$ instead of the average convergence
time, $t_{av}$,\cite{KS-LDPC} since we wish to prevent the dramatic
effect of a small fraction of finite samples with slow convergence or
no convergence.\cite{domany,median}

\begin{figure}
\vspace{0.5cm}
\centerline{\epsfxsize=2.5in \epsffile{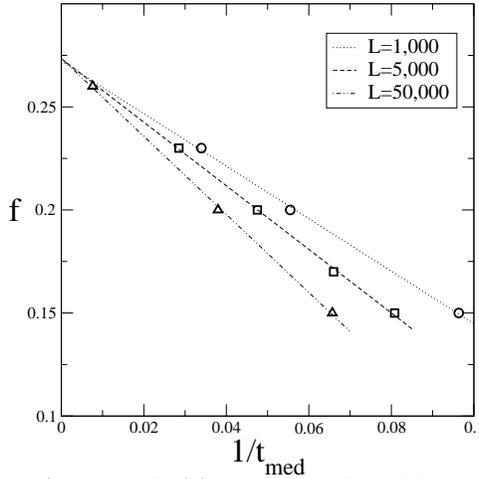}}
\caption{The flip rate $f$ as a function of $1/t_{med}$ for GF(4) with
$C_1=C_2=0.8$ and $L=1,000,~5,000~,50,000$.  The lines are a result of
a linear regression fit. The threshold, $f_c \sim 0.272$, extrapolated
from the scaling behavior eq. 7, is independent of $N$.  }
\end{figure}  

All simulation results presented below are derived for rate $1/3$ and
the construction of the matrices $A$ and $B$ of the MN code are taken
from \cite{KS-LDPC}. In all examined sets of autocorrelations, $10^3
\le L \le 5\!\times\!10^4$ and $4 \le q \le 64$, the scaling for the median
convergence time was indeed recovered.  At this stage an efficient
tool to estimate the threshold of an MN-DBP decoder exists and we are
ready to examine the efficiency of the DBP decoder as a function of
$\{C_k\}$ and $q$.

Results of simulations for $q=4,~8,~16$ and $32$ and selected sets of
autocorrelations are summarized in Table I, and the definition of the
symbols is: $\{C_k\}$ stand for the imposed values of autocorrelations
as defined in eqs.  1-2; $\{y_k\}$ are the interaction strengths,
eqs. 3 and 5; $H$ represents the entropy of sequences with the given
set of autocorrelations, eq. 2; $f_c$ is the estimated threshold of
the MN-DBP decoder derived from the scaling behavior of $t_{med}$;
$f_{Sh}$ is the Shannon's lower bound, eq. 4; Ratio is the efficiency
of our code $f_c/f_{Sh}$; $Z_R$ indicates the gzip compression rate
averaged over files of the sizes $10^6$ bits with the desired set of
autocorrelations. We assume that the compression rate with $L=10^6$
achieves its asymptotic ratio, as was indeed confirmed in the
compression of files with different $L$; $1/R^{\star}$ indicates the
ideal (minimal) ratio between the transmitted message and the source
signal after implementing the following two steps: compression of the
file using gzip and then using an ideal optimal encoder/decoder, for a given
BSC with $f_c$.  A number greater than (less than) $3$ in this column
indicates that the MN-DBP joint source-channel coding algorithm is
more efficient (less efficient) in comparison to the channel
separation method using the standard gzip compression.
The last four columns of Table I are devoted for the comparison of our
DBP algorithm to advanced compression methods. $PPM_R$ and $AC_R$
stand for the compression rate of files of the size $10^6$ bits with
the desired autocorrelations using the Prediction by Partial 
Match\cite{PPM} and for the Arithmetic Coder\cite{AC}, respectively. Similarly
to the gzip case, $1/R_{PPM}$ and $1/R_{AC}$ stand for the optimal
(minimal) rate required for the separation process (first a compression 
and then an ideal optimal encoder/decoder) assuming a BSC with $f_c$.

\begin{table}
\centerline{\epsfxsize=6.75in \epsffile{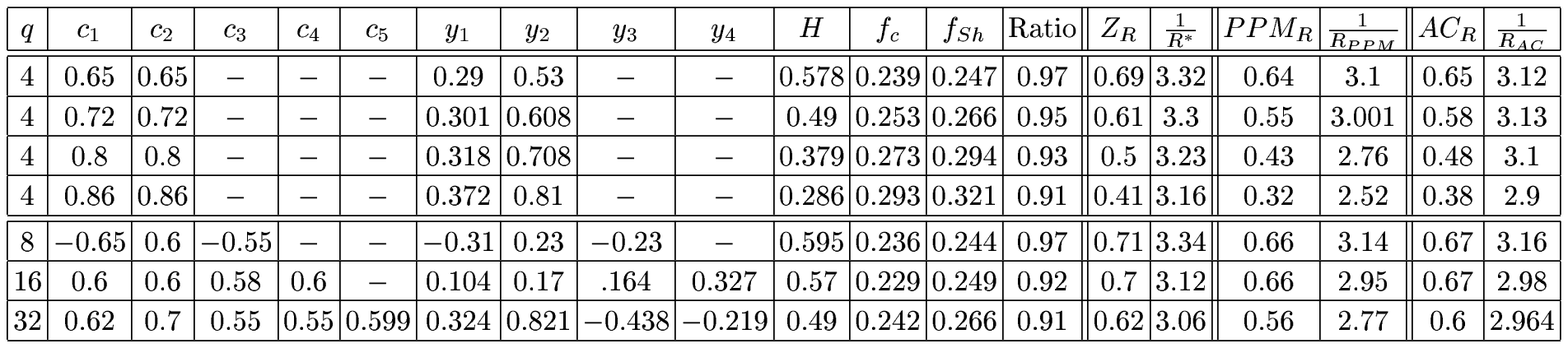}}
\caption{Results for $q=4,~8,~16,~32$ and selected sets of 
autocorrelations $\{C_k\}$.
}
\end{table}

Table I indicates the following main results: (a) For $q=4$ (the upper
part of Table I) a degradation in the performance is observed as the
correlations are enhanced and as a result the entropy decreases.  The
degradation seems to be significant as the entropy is below $\sim 0.3$
(or for the test case $R=1/3$, $f_c \ge 0.3$).\cite{bias} A similar
degradation was observed as the entropy decreases for larger values of
$q$. (b) The efficiency of our joint source-channel coding technique
is superior to the alternative standard gzip compression in the source
channel separation technique. For high entropy the gain of the MN-DBP
is about $5-10\%$.  This gain disappears as the entropy and the
performance of the DBP algorithm are decreased. (c) In comparison to
the standard gzip, the compression rate is improved by $2-5\%$ using
the AC method. A further improvement of a few percents is achieved by
the PPM compression. This later improvement seems to be significant in
the event of low entropy. (d) Our DBP joint source-channel coding
seems to be superior (by $\sim 3\%$)to the separation method based on
the PPM compression for high entropy. However for ensemble of
sequences characterized by low entropy this gain disappears.  (e) With
respect to the computational time of the source channel coding, our
limited experience indicates that the DBP joint-source channel coding
is faster than the AC separation method and the PPM separation method
is substantially slower.

For a given set of autocorrelations where $C_{k_0}$ is the maximal one
taken, the MN-DBP algorithm can be implemented with any field $q \ge
2^{k_0}$.  If one wishes to optimize the complexity of the decoder it
is clear that one has to work with the minimal allowed field,
$q=2^{k_0}$.  However, when the goal is to optimize the performance
of the code and to maximize the threshold, the selection of the
optimal field, $q$, is in question. In order to answer this question
we present in Fig. 2 results for $k_0=2$ ($C_1=C_2=0.86$) and
$q=4,~16,~64$. It is clear that the threshold, $f_c$, increases as a
function of $q$ as was previously found for the case of unbiased
signals.\cite{LDPC-GF(q)} More precisely, the estimated thresholds for
$q=4,~16,~64$ are $\sim 0.293,~0.3,~0.309$, respectively, and the
corresponding Ratios ($\equiv f_c/f_{Sh}$) are $0.913,~0.934, 0.962$.  where
$f_{Sh}=0.321$.  Note that the extrapolation of $f_c$ for large $q$
seems asymptotically to be consistent with $f_c(q) \sim 0.316
-0.18/q$.

\begin{figure}
\vspace{1.2cm}
\centerline{\epsfxsize=2.5in \epsffile{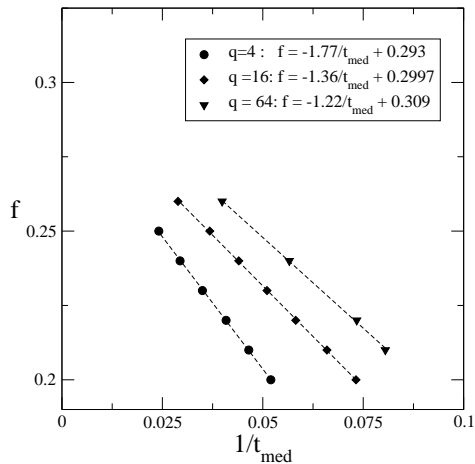}}
\caption{The scaling behavior, $f$ as a function of $1/t_{med}$, for
$C_1=C_2=0.86$ and $q=4,~16,~64$. The lines are a result of a linear
regression fit. The estimated thresholds for $q=4,~16,~64$ are
$0.293,~0.3,~0.309$, and the corresponding $Ratio \equiv f_c/f_{Sh}=
0.913,~0.934, 0.962$, where $f_{Sh}=0.321$.}
\end{figure}

For a given $q$, there are many sets of autocorrelations (or a finite
fraction of $\{C_k\}$ in $k_0-1$ dimensions) obeying the same entropy.
An interesting question is whether the performance of our DBP
algorithm measured by the Ratio $(\equiv f_c/f_{Sh})$ is a function of
the entropy only. Our numerical simulations indicate that the entropy
is not the only parameter which controls the performance of the DBP
algorithm. For the same entropy and $q$ the Ratio can fluctuate widely
among different sets of correlations.  For illustration, in Table II
results for two sets of autocorrelations with {\it the same entropy}
are summarized for each $q=4,~8,~16$ and $32$. It is clear that as the
Ratio $(\equiv f_c/f_{Sh})$ is much degradated the gzip performance is
superior (the second example with $q=8$ and $32$ in Table II where the
Ratio is $0.8$ and $0.72$, respectively).  The crucial question is to
find the criterion to classify the performance of the DBP algorithm
among all sets of autocorrelations obeying the same entropy.  Our
generic criterion is {\it the decay of the correlation function over
distances beyond two successive blocks}.  However, before the
examination of this criterion, we would like to turn back to some
aspects of statistical physics.

The entropy of sequences with given first $k_0$ correlations are
determined via the effective Hamiltonian consisting of $k_0$
interactions, eqs. 2-3.  As a result the entropy of these sequences is
{\it the same} as the entropy of the effective Hamiltonian,
$H\{y_k\}$, at the inverse temperature $\beta=1$, eq. 5.  As for the
usual scenario of the transfer matrix method, the leading order of
quantities such as the free energy and the entropy are a function of
the {\it largest eigenvalue} of the transfer matrix only. On the other
hand the decay of the correlation function is a function of the whole
spectrum of the $2^{k_0}$ eigenvalues.  Asymptotically, the decay of
the correlation function is determined from the ratio between the
second largest eigenvalue and the largest eigenvalue,
$\lambda_2/\lambda_{max}$.  From the statistical mechanical point of
view one may wonder, why the first $k_0$ correlations can be
determined using the information of $\lambda_{max}$ only. The answer
to this question is that once the transfer matrix is defined as a
function of $\{y_k\}$, eqs. 3-5, {\it all eigenvalues} are determined
as well as $\lambda_{max}$. There is no way to determine
$\lambda_{max}$ independently of all other eigenvalues.

In Table II results of the DBP-MN algorithm for $q=4,~8,~16,~32$ are
presented. For each $q$, two different sets of autocorrelations
characterized by the {\it same entropy} and threshold $f_{Sh}$ are
examined.  The practical method we used to generate different sets of
autocorrelations with the same entropy was a simple Monte Carlo over
the space of $\{C_k\}$.\cite{haggai1} The additional column in Table
II (in comparison with Table I) is the ratio between
$\lambda_2/\lambda_{max}$, which characterizes the decay of the
correlation function over large distances. Independent of $q$, it is
clear that for a given entropy as $\lambda_2/\lambda_{max}$
increases/decreases the performance of the DBP algorithm measured by
the Ratio $f_c/f_{Sh}$ is degradated/enhanced.  The new criterion to
classify the performance of the DBP algorithm among all sets of
autocorrelations obeying the same entropy is the decay of the
correlation function.  This criterion is consistent with the tendency
that as the first $k_0$ autocorrelations are increased/decreased a
degradation/enhancement in the performance is observed (see Table I).
The physical intuition is that as the correlation length increases,
the relaxation time increases and flips on larger scales than nearest
neighbor blocks are required.

Note that the decay of the correlation function in the intermediate
region of a small number of blocks is a function of all the $2^{k_0}$
eigenvalues.  Hence, in order to enhance the effect of the fast decay
of the correlation function in the case of small
$\lambda_2/\lambda_{max}$, we also try to enforce in our Monte Carlo
search that all other $2^{k_0}-2$ eigenvalues be less than
$A\lambda_{max}$ with the minimal possible constant $A$.  This
additional constraint was easily fulfilled for $q=4$ with $A=0.1$, but
for $q=32$ the minimal $A$ was around $0.5$.

Finally we raise the question of whether for a given entropy a
degradation in the {\it typical} performance of the DBP algorithm is
expected as $q$ increases.  This is crucial since the superiority, if
any, of the DBP joint source-channel coding method over advanced
compression methods is in question. As explained above, our Monte
Carlo simulations indicate that for a given entropy the suppression of
the correlation function is more difficult as $q$
increases.\cite{haggai1} This is a strong indication that as $q$
increases a degradation in the typical performance of the DBP decoder
is expected, but its nature and significance have still to be examined
in further research.

We thank Shlomo Shamai and David MacKay for valuable discussions.

\begin{table}
\centerline{\epsfxsize=6.75in \epsffile{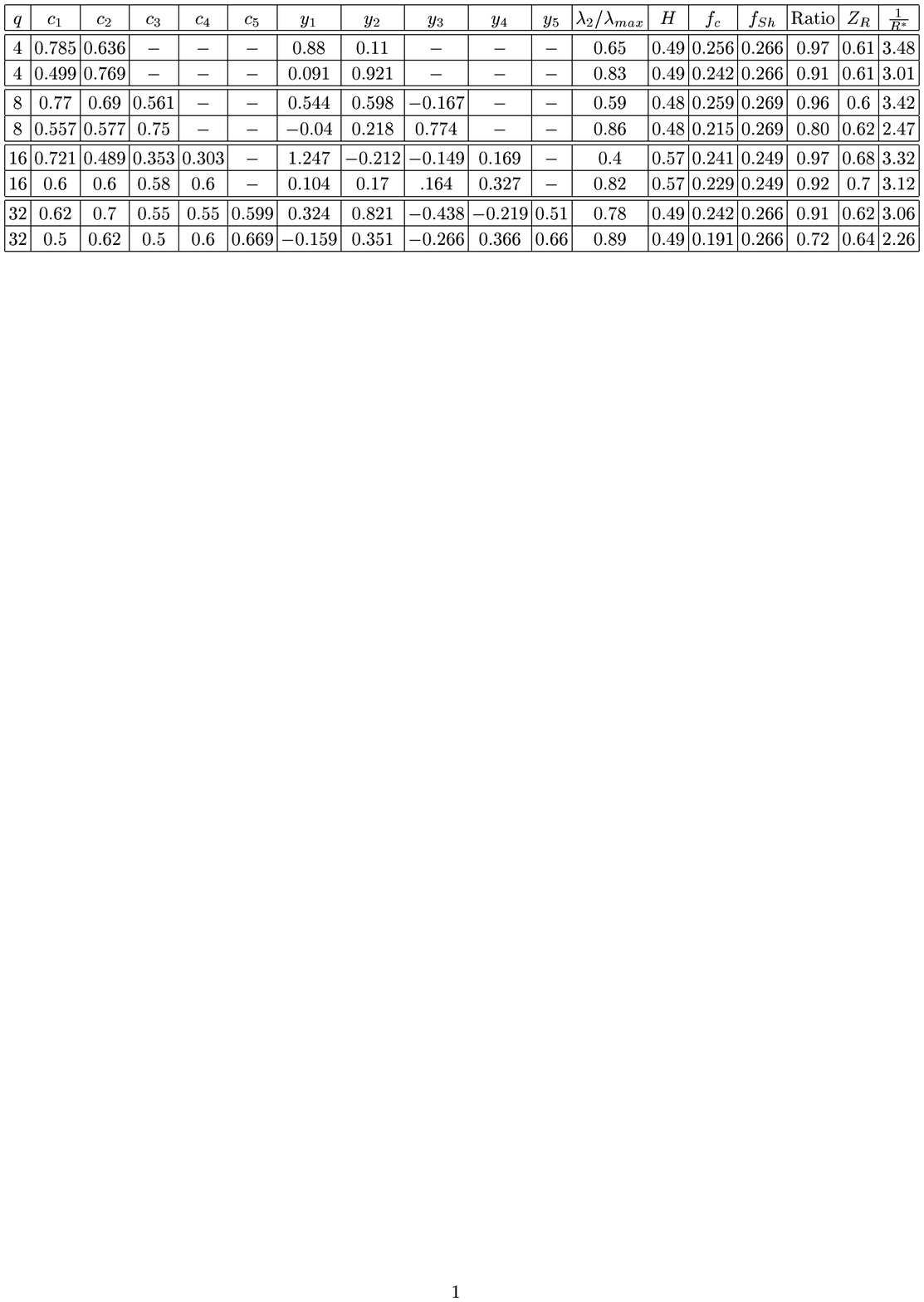}}
\caption{Results for $q=4,~8,~16,~32$ and different sets of
autocorrelations.  For each $q$, two different sets of
autocorrelations characterized by the same entropy and threshold
$f_{Sh}$ are examined.  As $\lambda_2/\lambda_{max}$
increases/decreases the performance of the DBP algorithm measured by
the Ratio $f_c/f_{Sh}$ is degradated/enhanced.
}
\end{table}




\vskip-12pt
\begin{thebibliography}{99}


\bibitem{shamail1} J. Kliewer and R. Thobaben, "Combining FEC and
Optimal Soft-Input Source Decoding for Reliable Transmission of
Correlated Variable-Length Encoded Signals" (to appear in DCC2002)

\bibitem{shamail2} S. Shamai and S. Verdu, \emph{European Transactions on
Telecommunications}, vol. 6, no. 5, pp. 587-600, Sep.-Oct. 1995

\bibitem{shamail3}A.D. Liveris, Z. Xiong and C.N. Georghiades,
"Compression of Binary Sources with Side Information Using Low-Density
Parity-Check Codes", paper CTS-08-7, GLOBECOM2002, 17-21, November, 
Taipei, Taiwan

\bibitem{shamail4} J. Garcia-Frias and J.D. Villasenor, 
\emph{IEEE J. Selec. Areas Commun.}, Vol. 19, No. 9, pp. 1671-1679, Sept. 2001

\bibitem{shamail5} C.-C. Zhu and F. Alajaji, \emph{IEEE
Commun. Letters}, Vol. 6, No. 2, pp. 64-66, February 2002

\bibitem{shamail6} J. Garcia-Frias and Z. Ying, \emph{IEEE Commun.
Letters}, Vol. 6, No. 9, pp. 394-396, September 2002

\bibitem{Shannon-48} C. E. Shannon. A mathematical theory of
communication. \emph{Bell System Technical J.}, {\bf 27}, 379-423,
623-656, 1948.

\bibitem{ido-new} I. Kanter and H. Rosemarin, cond-mat-0301005

\bibitem{sourlas} N. Sourlas, \emph{Nature} \textbf{339}, 6227, 1989

\bibitem{sourlas1} N. Sourlas, \emph{Physica A} \textbf{302}, 14 (2001)

\bibitem{baxter}R. J. Baxter. Exactly Solved Models In Statistical
Mechanics. \emph{Academic Press}, 1982

\bibitem{LDPC-GF(q)}M. C. Davey and D. J. C. Mackay, \emph{IEEE
Communications Letters}, Vol 2, No. 6, June 1998


\bibitem{David_Mackay} We thank David MacKay for bringing to our attention
the reduced complexity of the decoder from $O(q^2)$ to $O(qlogq)$

\bibitem{David_Mackay1} D. J. C.  MacKay and M. C. Davey, "Gallager
Codes for Short Block Length and High Rate Applications", Codes,
Systems and Graphical Models", IMA Volumes in Mathematics and its
Applications, Springer-Verlag (2000).

\bibitem{David_Mackay2} T. Richardson and R. Urbanke,  
\emph{IEEE Transactions on Information Theory}, {\bf 47}, 599 (2001).

\bibitem{KS-LDPC}I. Kanter and D. Saad. \emph{Phys. Rev. Lett.}
\textbf{83}, 2660, 1999

\bibitem{KS-Gaussian}I. Kanter and D. Saad. 
\emph{ J. Phys. A} \textbf{33}, 1675 (2000)

\bibitem{median} In practice we define $t_{med}$ to be the average
convergence time of all samples with $t \le$ the median time.

\bibitem{domany} A. Priel, M. Blatt, T. Grossman, E. Domany and I. Kanter
{\sl Phys. Rev. E} {\bf 50,} 577 (1994)



\bibitem{PPM} The used PPMZ software can be downloaded from
www.cbloom.com/src/ppmz.html


\bibitem{AC} The used AC software can be downloaded from
www.cs.mu.oz.au/~alistair/arith\_coder


\bibitem{bias} A similar degradation in the performance was observed
for $q=2$ and biased binary messages (each source bit is equal $0/1$ and is
chosen with probability $p/1-p$).  As $|p-0.5|$ increases the
entropy decreases and a degradation in the performance of the MN
algorithm was observed.


\bibitem{haggai1} H. Kfir and I. Kanter (in preparation)

\end{thebibliography}
 \end{document}